\documentclass{PoS}

\usepackage{epsfig}

\title{Non-lattice simulation of supersymmetric gauge theories as a probe to quantum black holes and strings}
\ShortTitle{Non-lattice simulation of supersymmetric gauge theories...}
\author{\speaker{Jun Nishimura}%
\\
        KEK Theory Center,
Tsukuba 305-0801, Japan,\\ 
and 
Graduate University for Advanced Studies (SOKENDAI), 
Tsukuba 305-0801, Japan\\
        E-mail: \email{jnishi@post.kek.jp}}
\abstract{In the past decade we have witnessed remarkable developments 
in the gauge-gravity duality, which suggested a new approach
to superstring theory and quantum space-time. In this context
it is important to study supersymmetric large-$N$ gauge theories
in the strongly coupled regime. I will summarize the results
and insights obtained so far by non-lattice simulations.
A simple example of the gauge-gravity duality is 
the one between 1d U($N$) gauge theory with 16 supercharges and 
the so-called black 0-brane solution in type IIA supergravity.
In order for this duality to be valid, one has to take the 
't Hooft large-$N$ limit and to take the strong coupling limit 
on the gauge theory side.
The gauge theory can be regularized by fixing the gauge completely
thanks to one dimension, and by introducing a Fourier mode cutoff.
One can then use the standard RHMC algorithm to simulate the system.
The energy calculated as a function of the temperature was compared 
with the results obtained from the gravity side based on the black 
hole thermodynamics. This confirmed the gauge-gravity duality with 
high accuracy and provided the microscopic origin of the black hole 
thermodynamics. From the calculation of the Wilson loop, one obtains 
the Schwarzschild radius of the dual geometry.
One can actually use the present 1d model with supersymmetric 
mass deformation to study ${\cal N}=4$ super Yang-Mills theory 
on $R \times S^3$ based on a novel large-$N$ reduction, which generalizes 
the original idea of Eguchi and Kawai.
A test of this approach has been provided by Monte Carlo simulation at
weak coupling. It is remarkable that we can now simulate the 
4d superconformal field theory, which appears in the most typical case 
of the gauge-gravity duality known as the AdS/CFT correspondence. 
In particular, no fine-tuning is required unlike previous proposals 
based on the lattice regularization.}

\FullConference{The XXVII International Symposium 
on Lattice Field Theory - LAT2009\\
		 July 26-31 2009\\
		 Peking University, Beijing, China}

\newcommand {\beq} {\begin{equation}}
\newcommand {\eeq} {\end{equation}}
\newcommand {\beqa}{\begin{eqnarray}}
\newcommand {\eeqa}{\end{eqnarray}}

\newcommand {\tr}{{\rm tr\,}}

\newcommand {\ee}{\mbox{e}}

\begin{document}

\section{Introduction}

I hesitate a bit to talk about \emph{non-lattice} simulation at
this \emph{lattice} conference, but this is indeed crucial 
for the purpose
of simulating supersymmetric theories.
What I am going to discuss is the so-called gauge-gravity duality,
which is a conjecture from superstring 
theory \cite{AdS-CFT,Gubser:1998bc}.
%,Witten:1998qj}.
(See ref.\ \cite{Aharony:1999ti} for a comprehensive review.)
The statement itself is simple, and we don't even have to refer to
superstring theory.
Let us consider U($N$) supersymmetric Yang-Mills theory (SYM) with 
16 supercharges (or 32 supercharges, in a special case).
We take the so-called 't Hooft limit, which amounts to sending 
$N$ to infinity with fixed $\lambda\equiv g_{\rm YM}^2 N$.
Next we consider the strongly coupled regime, 
namely the large-$\lambda$ regime.
Then the statement  is that the SYM
%supersymmetric Yang-Mills theory
is ``dual'' to 
%has a dual description in terms of 
a classical solution in 10d supergravity.

\begin{figure}[htb]
\begin{center}
\includegraphics[height=6cm]{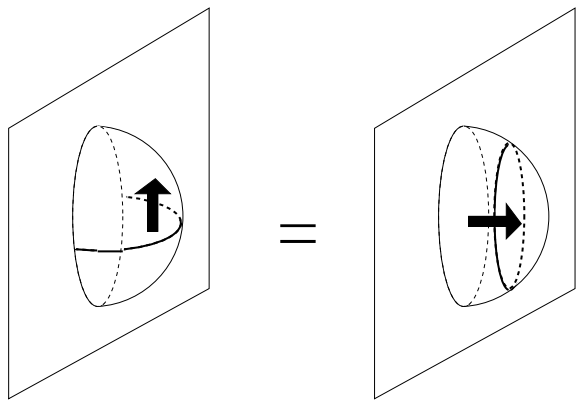}
%{bfss_continuum.eps}
\end{center}
\caption{
%The absolute value of
On the left, an open string attached to the D brane
is propagating along it.
On the right, the same process has been viewed as
emission of a closed string from the D brane.
}
\label{dbrane}
\end{figure}

The argument for this conjecture is actually very intuitive and easy to
understand.
In superstring theories, there exists
a soliton-like object termed D brane.
``D'' stands for the Dirichlet boundary condition
imposed at the boundary of the worldsheet of a string.
D brane can extend in $p+1$ dimensions, and it is characterized
as a hypersurface on which strings can end on.
Let us consider an open string attached to the D brane
propagating along it.
In figure \ref{dbrane} on the left, 
we describe such a process diagrammatically.
If one slices the diagram in the orthogonal direction,
one notices that the same process can be viewed as emission of
a closed string.
This is an example of the well-known notion of open-string/closed-string
duality.
Note here that an open string includes a gauge particle as a massless mode,
and similarly a closed string includes a graviton.
Let us consider $N$ D branes lying on top of each other.
Then, in the low energy limit, 
one obtains $(p+1)$-dimensional U($N$) SYM
%super Yang-Mills theory 
as an effective theory which describes the massless degrees
of freedom of open strings attached to the D branes.
On the other hand, one obtains a curved 10d space-time in the bulk
since the D brane sources gravitons.
In order for the supergravity to be valid as a low energy
and classical description of superstring theory in the bulk,
one has to take the 
so-called 't Hooft large-$N$ limit with fixed $\lambda\equiv g_{\rm YM}^2 N$,
and then to take the large-$\lambda$ limit.
This is so, since the string loop corrections are suppressed by $1/N$,
whereas the $\alpha '$ corrections, which are 
due to strings having finite extent,
are suppressed by some powers of $1/\lambda$.

%%%%%%%%%%%%%%%%%%

Why is this duality interesting?
First of all, it is a realization of an old idea 
by 't Hooft \cite{'tHooft:1973jz},
which states that the large-$N$ gauge theory is equivalent to 
some classical string theory, although in those days people may not
have anticipated that the string theory actually lives in a curved 
space-time.
It is interesting that the curved space-time
emerges from a gauge theory in a flat space.
This aspect of the duality is often referred to as 
the \emph{emergent space-time}.
In the gauge-gravity duality, one typically obtains 
%(asymptotically)
the anti-de Sitter space. 
If one considers the gauge theory at finite temperature,
one obtains a black-hole-like geometry \cite{Witten:1998zw,Itzhaki:1998dd}.
As applications, one can study strongly coupled gauge theories,
which are relevant to hadron and condensed matter physics,
from a curved space-time.
One can also use the duality in the opposite direction,
and try to explain the microscopic origin of the black hole thermodynamics
in terms of gauge theory.
An ultimate goal of the gauge-gravity duality is to construct
a non-perturbative and background independent formulation 
of superstring theory by using gauge-theory degrees of freedom.

Since the gauge-gravity duality is a strong-weak duality,
it is important to study gauge theories in the strongly coupled regime. 
Monte Carlo simulation can be a powerful tool for such purposes.
However, the problem is that the gauge theories we are interested in
have supersymmetry, which is broken by the lattice.
This can be seen immediately if one recalls the supersymmetry algebra
$\{Q , \bar{Q} \} \propto P_\mu $, where the generators for translation
appear on the right hand side. Since the translational symmetry is broken
by the lattice regularization, one necessarily breaks supersymmetry.
%One can still hope to restore supersymmetry in the continuum limit,
%but that typically requires fine-tuning.

Recently there are considerable developments 
in ``lattice supersymmetry'',
which can be categorized into two classes.
One is the construction of lattice actions with various symmetries.
For instance, one can preserve one supercharge by using the so-called
topological twist. (See Catterall's contribution of this volume).
The other one, which we discuss here, is
non-lattice simulations \cite{Hanada-Nishimura-Takeuchi}
of supersymmetric 
gauge theories in 1 dimension 
with 16 supercharges \cite{AHNT,Hanada:2008gy,Hanada:2008ez}.
Notably, one can extend this approach 
to 3d and 4d gauge theories \cite{Ishiki:2008te}
%,Ishiki:2009sg}
by using the idea of large-$N$ reduction \cite{Ishii:2008ib}.
In the 4d case, the gauge theory becomes superconformal and the 
number of supersymmetries enhances from 16 to 32.
This superconformal theory is interesting on its own right,
but it is also studied intensively in the context of 
the AdS/CFT correspondence, which is a typical case of the
gauge-gravity duality \cite{AdS-CFT}.
The non-lattice simulation of the 4d superconformal theory
requires no fine-tuning,
unlike the previous proposals based on the 
lattice regularization \cite{latticeSUSY_N4}.

This article is organized as follows.
In section \ref{sec:non-lattice} I discuss the non-lattice simulation of 1d SYM
with 16 supercharges.
In particular, I explain how black hole thermodynamics appear
from 1d SYM, and how the Schwarzschild radius appears from the Wilson loop.
In section \ref{sec:higherD} I review the large-$N$ reduction,
which enables us to extend these works to higher dimensions.
% based on the large-$N$ reduction.
In particular, I discuss how one can study $\mathcal{N}=4$ SYM
on $R \times S^3$ in the 't Hooft limit, and present some 
preliminary results for the Wilson loop and the two-point 
correlation functions.
In section \ref{sec:summary} I conclude with a summary.
% and some future prospects.

\section{Non-lattice simulation of 1d SYM with 16 supercharges}
\label{sec:non-lattice}

The 1d SYM with 16 supercharges has the following actions
for the bosonic part and the fermionic part, respectively.
\beqa
S_{\rm b} &=&  \frac{1}{g^2} \ \int_{0} ^{\beta}  dt \
\tr \left\{  \frac{1}{2} \Bigl(D X_i(t) \Bigr)^2
- \frac{1}{4} [X_i(t) , X_j (t)] ^2
\right\}  \\
S_{\rm f}&=&  \frac{1}{g^2} \ \int_{0} ^{\beta}  dt \
\tr \left\{  \frac{1}{2}\Psi_\alpha D \Psi _\alpha
- \frac{1}{2} \Psi_\alpha (\gamma_i)_{\alpha\beta}
 [X_i , \Psi_\beta]  \right\}
\label{bfss-action}
\eeqa
It is a 1d U($N$)
gauge theory, and the covariant derivative is denoted as
$D = \partial_t  - i \,  [A (t), \ \cdot \ ]$.
$X_j(t) \ (j=1, \cdots , 9)$
and $\Psi_\alpha(t) \ (\alpha=1, \cdots , 16) $
are $N\times N$ Hermitian matrices, and the theory has SO(9) symmetry.
When we are interested in finite temperature,
we impose periodic boundary conditions on $X_j(t)$ 
and anti-periodic boundary conditions on $\Psi_\alpha(t)$.
Then the temperature is given by $T\equiv \beta^{-1}$,
where $\beta$ is the extent in the Euclidean time ($t$) direction.
The 't Hooft coupling constant is defined by
$\lambda \equiv g^2 N$, which has the dimension of mass cubed.
The physics of the system is determined only by the dimensionless 
coupling constant $\lambda_{\rm eff} \equiv \frac{\lambda}{T^3}$.
Therefore one can take $\lambda=1$ without loss of generality.
With this convention, the low $T$ regime corresponds to the 
strongly coupled regime, which is expected to have the dual gravity
description \cite{Itzhaki:1998dd}, 
whereas the high $T$ regime is essentially weakly coupled,
and the high temperature expansion (HTE) is applicable \cite{HTE}.

In non-lattice simulation \cite{Hanada-Nishimura-Takeuchi},
we introduce
an upper bound on the Fourier mode as
$X_i  (t) = \sum_{n=-\Lambda}^{\Lambda} 
\tilde{X}_{i ,n} \ee^{i \omega n t}$, where
$\omega = \frac{2\pi}{\beta}$,
and similarly for the fermions.
This idea does not work usually because it breaks gauge invariance.
(Recall that the Fourier mode is not a gauge invariant concept.)
However, in 1d, one can fix the gauge non-perturbatively in the 
following way.
We first take the static diagonal gauge
$A(t) = 
\frac{1}{\beta}
{\rm diag}(\alpha_1 , \cdots ,
\alpha_N )$, in which the gauge field
is constant in time and diagonal.
By following the usual Faddeev-Popov procedure one obtains 
\beq
S_{\rm FP}
= - \sum_{a<b}
2 \ln  \left|\sin \frac{\alpha_a -\alpha_b}{2} \right|
\eeq
as a term to be added to the action.
The above gauge choice does not fix the gauge symmetry completely,
and there is a residual symmetry given by
\beq
\alpha_a \mapsto  \alpha_a + 2 \pi \nu_a \ , 
\quad \quad
\tilde{X}_{i, n}^{ab} \mapsto  \tilde{X}_{i , n-\nu_a + \nu_b}^{ab} \ , 
\quad \quad 
\tilde{\Psi}_{\alpha, n}^{ab} 
\mapsto  \tilde{\Psi}_{\alpha , n-\nu_a + \nu_b}^{ab} \ ,
\eeq
which represents a topologically nontrivial gauge transformation
corresponding to the gauge function
$g(t) = {\rm diag} (\ee ^{i \omega \nu_1 t } , \cdots ,
\ee ^{i \omega \nu_N t } )$.
This residual gauge symmetry can be fixed by imposing
$-\pi < \alpha_a \le \pi$.
One can then introduce the Fourier mode cutoff $\Lambda$.
%This breaks supersymmetry, but it restores very
%quickly as one increases $\Lambda$.
%Also, since 
Since there is no UV divergence in this 1d model,
one can take the $\Lambda \rightarrow \infty$ limit naively,
and one obtains the original
gauge theory with 16 supercharges.

The system with finite $\Lambda$
%after all these procedures
% composed of finite degrees of freedom, which 
can be simulated efficiently by using
the standard RHMC algorithm \cite{Clark:2003na}.
%\cite{Clark:2004cp}.
In particular, the Fourier acceleration \cite{Catterall:2001jg}
can be implemented without
extra cost since we are dealing with the Fourier modes directly as
the fundamental degrees of freedom. This is crucial in reducing the
critical slowing down 
at large $\Lambda$.
%at low temperature.
% due to the fact that the theory contains
%massless bosons and fermions, although the coupling constant is 
%dimensionful.
(The same theory is also studied using the standard lattice
approach \cite{Catterall:2007fp}.
%{Catterall:2008yz}.
%since there is no relevant operator which breaks supersymmetry
%in this theory.
However, from the results obtained so far,
the non-lattice simulations seem to be far more
efficient in obtaining the continuum limit.)

Let us first discuss the phase structure that appears
when one changes the temperature.
As is well known, the Polyakov line serves as an order parameter
for the spontaneous breaking of the center symmetry.
Figure \ref{fig:old} (Left) shows the results \cite{AHNT}.
At high temperature the data agree nicely with 
%the high temperature expansion \cite{HTE} 
the HTE \cite{HTE} 
including the next-leading order.
As the temperature decreases below $T \sim 3$, the data start to
deviate, and at low temperature below $T \sim 0.9$, the data can be
fitted to the characteristic behavior of the ``deconfined phase''
\beq
\langle |P| \rangle = \exp \left(-\frac{a}{T}+b \right) \ .
\label{deconf-P}
\eeq
In the temperature regime investigated, we find no phase transition.
This is in sharp contrast to the bosonic model
\cite{latticeBFSS,Aharony:2004ig,Kawahara:2007fn},
%,Mandal:2009vz,Kawahara:2007nw},
which undergoes
a phase transition to the ``confined phase''
at $T \sim 0.9$.
The absence of the phase transition is consistent
with analyses on the gravity 
side \cite{Witten:1998zw,Barbon:1998cr}.
%,Aharony4}.

\FIGURE{
    \epsfig{file=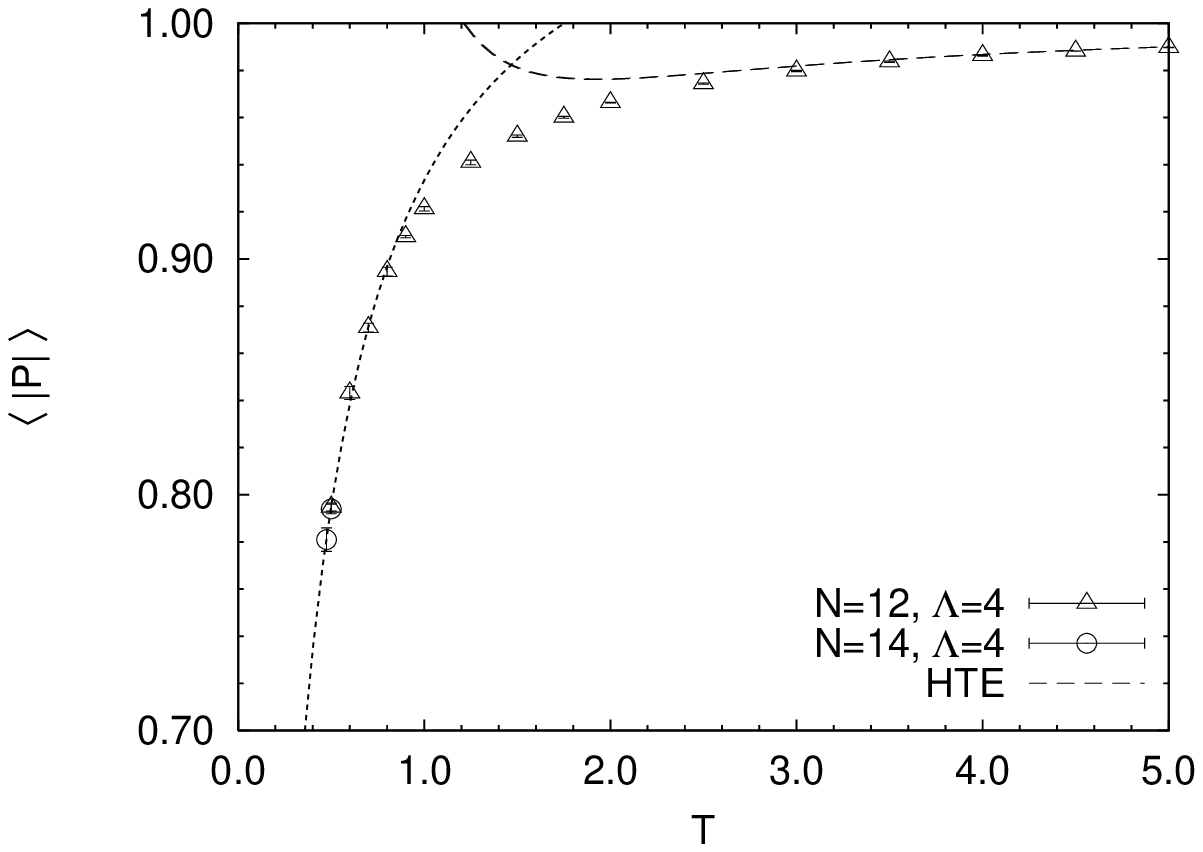,%
%angle=270,
width=7.0cm}
    \epsfig{file=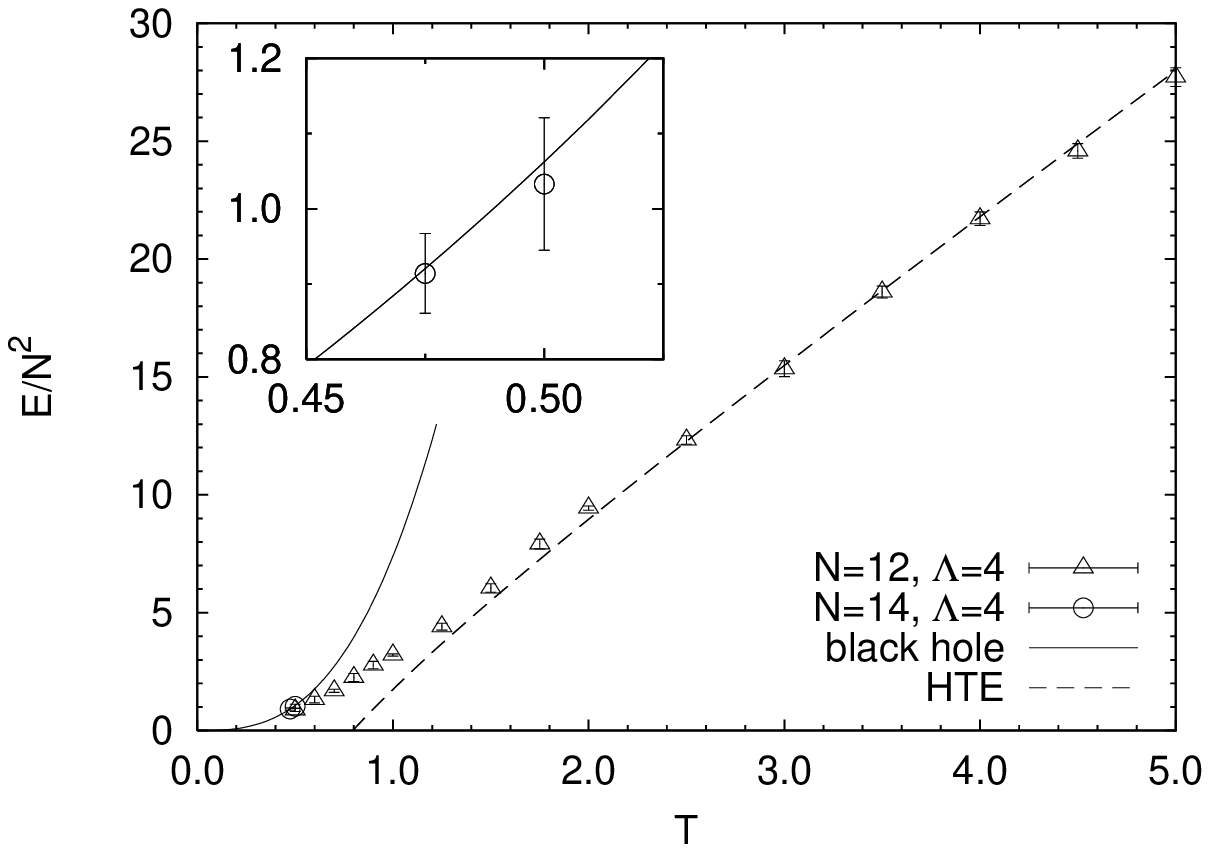,%
%angle=270,
width=7.0cm}
\caption{(Left) 
The Polyakov line
is plotted against 
$T$.
The dashed line represents the result
%obtained by 
of HTE
up to the next leading order for $N=12$ \cite{HTE}.
The dotted line represents a fit to
eq.\ (\protect\ref{deconf-P})
with $a=0.15$ and $b=0.072$.
(Right) 
The energy
(normalized by $N^2$)
is plotted against $T$.
The dashed line represents the result
obtained by HTE
up to the next leading order 
for $N=12$ \cite{HTE}.
The solid line represents
the asymptotic
power-law behavior at small $T$
predicted by the gauge-gravity duality.
The upper left panel zooms
up the region, where
the power-law behavior sets in.
}
\label{fig:old}
}
%%%%%%%%%%%%%%%%%%

Let us turn to a quantitative prediction from the gauge-gravity
duality.
Given the dual geometry, one can use Hawking's theory 
of the black hole thermodynamics to obtain various thermodynamic
relations such as \cite{Klebanov:1996un}
\beq
\frac{1}{N^2}\left( \frac{E}{\lambda^{1/3}}
\right)= c
 \, \left( \frac{T}{\lambda^{1/3}} \right)
^{14/5}  \ , \quad \quad
c =  \frac{9}{14}
\left\{ 4^{13} 15^{2} \left ( 
\frac{\pi}{7}
\right) ^{14}
 \right\} ^{1/5} = 7.41 \cdots \ .
\label{gravity-leading}
\eeq
%% where the coefficient $c$ is given by
%% \beq
%% c =  \frac{9}{14}
%% \left\{ 4^{13} 15^{2} \left ( 
%% \frac{\pi}{7}
%% \right) ^{14}
%%  \right\} ^{1/5} = 7.41 \cdots \ .
%% \eeq
The gauge-gravity duality
predicts that this should be reproduced
by 1d SYM in the large-$N$ limit at low $T$ \cite{Itzhaki:1998dd}.
The importance of this prediction is that,
if it is true,
it explains the microscopic origin of the black hole thermodynamics,
meaning that the 1d SYM provides the quantum description of 
the states inside the black hole. 

In figure \ref{fig:old} (Right)
we plot the internal energy \cite{AHNT}, which is defined by
$E = \frac{\partial}{\partial \beta} (\beta {\cal F})$
in terms of the free energy $\mathcal{F}$.
At $T\gtrsim 3$
the data agree with 
%the high temperature expansion \cite{HTE}.
the HTE \cite{HTE}. 
%and start to deviate at 
As one goes to lower temperature,
the data points approach the solid line, which 
represents the result (\ref{gravity-leading})
obtained from the 10d black hole.
(See refs.\ \cite{KLL} for earlier studies based on
the Gaussian approximation.)
%which is consistent with the prediction from the gauge-gravity duality.

The plots in figure \ref{fig:old}
were actually presented two years ago 
at LATTICE 2007 in Regensburg \cite{Nishimura:2008ta}.
A common criticism in those days was that 
it was not clear whether the gauge theory
results continue to follow the line predicted from gravity at lower $T$.
In fact, simulations at lower $T$ are difficult, since one has to
increase $\Lambda$ proportionally to $1/T$, and at the same time
one has to increase $N$ to avoid the run-away behavior 
due to finite $N$ \cite{AHNT}.
%, and the thermal mass of fermions decreases.
Instead of lowering $T$, 
%we took a cleverer path.
%In two years we have accumulated more data 
%with larger $N$ and larger $\Lambda$.
%Namely, 
we were able to determine
the power of the subleading term as \cite{Hanada:2008ez}
\beq
\frac{1}{N^2}\left( \frac{E}{\lambda^{1/3}}
\right)= c
 \, \left( \frac{T}{\lambda^{1/3}} \right)
^{14/5}  
- C \,  \left( \frac{T}{\lambda^{1/3}} \right)
^{23/5} 
\ ,
%% \frac{E}{N^2}
%% =7.41 \, 
%%  T^{14/5} - C \,  T^{23/5} 
\label{sub-leading}
\eeq
from gravity.
%The subleading term is due to 
This was derived by considering
higher derivative corrections
in the supergravity action due to the effects
of strings having finite extent ($\alpha ' $ corrections).
The coefficient $C$ of the subleading term
is calculable in principle, but it requires 
the full information of the higher derivative corrections, which are
yet to be determined.
By using (\ref{sub-leading}), however,
we can already make a nontrivial test of 
the gauge-gravity duality \cite{Hanada:2008ez}.
In figure \ref{fig:energy} (Left) we plot the discrepancy
$7.41 T^{14/5}-E/N^2$ against $T$ in the log-log scale,
which reveals that the power of the subleading term is indeed 
consistent with the predicted value $23/5=4.6$.
In figure \ref{fig:energy} (Right) we find that the data at $T\lesssim 0.7$ 
can be nicely fitted to the form (\ref{sub-leading}) with $C=5.58$.
Note also that the $\Lambda=6$ data seem to suffer 
from some finite $\Lambda$ effects at low $T$. 
From this point of view, we consider that 
the $\Lambda=4$ data points at low $T$
% with $\Lambda=4$
in figure \ref{fig:old} (Right),
which seem to be on the curve of the leading order 
result from gravity, also suffer from finite $\Lambda$ effects.
%We actually need the subleading term for precise agreement.
Now we know that actually the subleading term 
in (\ref{sub-leading})
should be taken into account 
for precise agreement.
%is actually needed for precise agreement.

\FIGURE{
    \epsfig{file=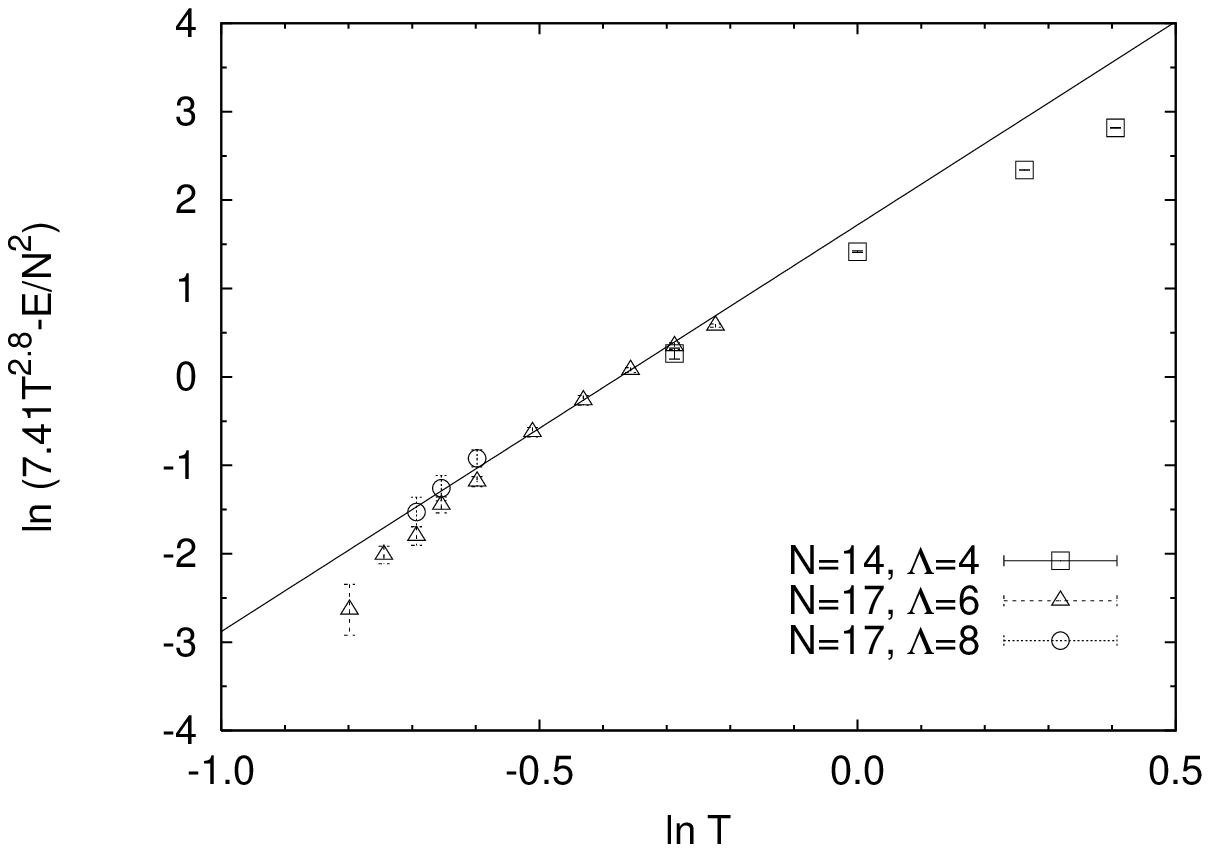,%
%angle=270,
width=7.0cm}
    \epsfig{file=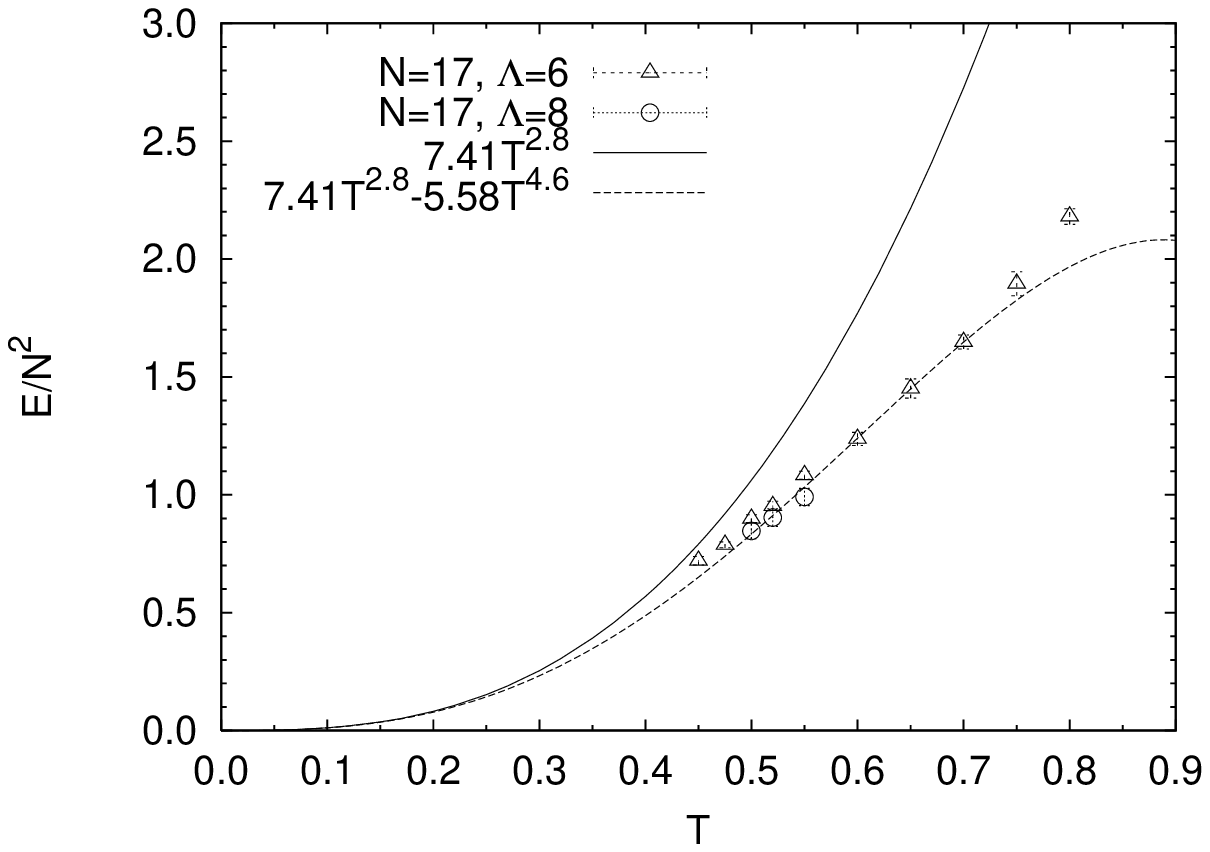,%
%angle=270,
width=7.0cm}
\caption{(Left) 
The deviation of the internal energy
$\frac{1}{N^2} E$ from the leading term
$7.41 \, T^{\frac{14}{5}}$
%predicted from the gravity
is plotted against the temperature
in the log-log scale for $\lambda=1$.
The solid line represents 
a fit to a straight line with the slope 4.6
predicted from the $\alpha '$ corrections 
on the gravity side.
(Right) 
The internal energy $\frac{1}{N^2} E$
%(normalized by $N^2$)
is plotted against $T$ for $\lambda=1$.
%versus 
%temperature.
%for 
%$N=8$, $\Lambda=2$ at $T\ge 1$ and 
%$N=12$, $\Lambda=4$ at $T\le 1$.
The solid line represents
the leading asymptotic behavior at small $T$
predicted by the gauge-gravity duality.
%based on the results obtained from 
%the dual black-hole geometry.
The dashed line represents
a fit to the behavior (\ref{sub-leading})
including the subleading term
%the behavior including the subleading term
with $C=5.58$.
%% The dashed line represents the result
%% obtained by the high $T$ expansion
%% up to the next leading order 
%% for $N=12$ \cite{HTE}.
}
\label{fig:energy}
}

%%%%%%%%%%%%%%%%%%%%%%%%%%%%%%

\begin{figure}[htb]
\begin{center}
\includegraphics[height=6cm]{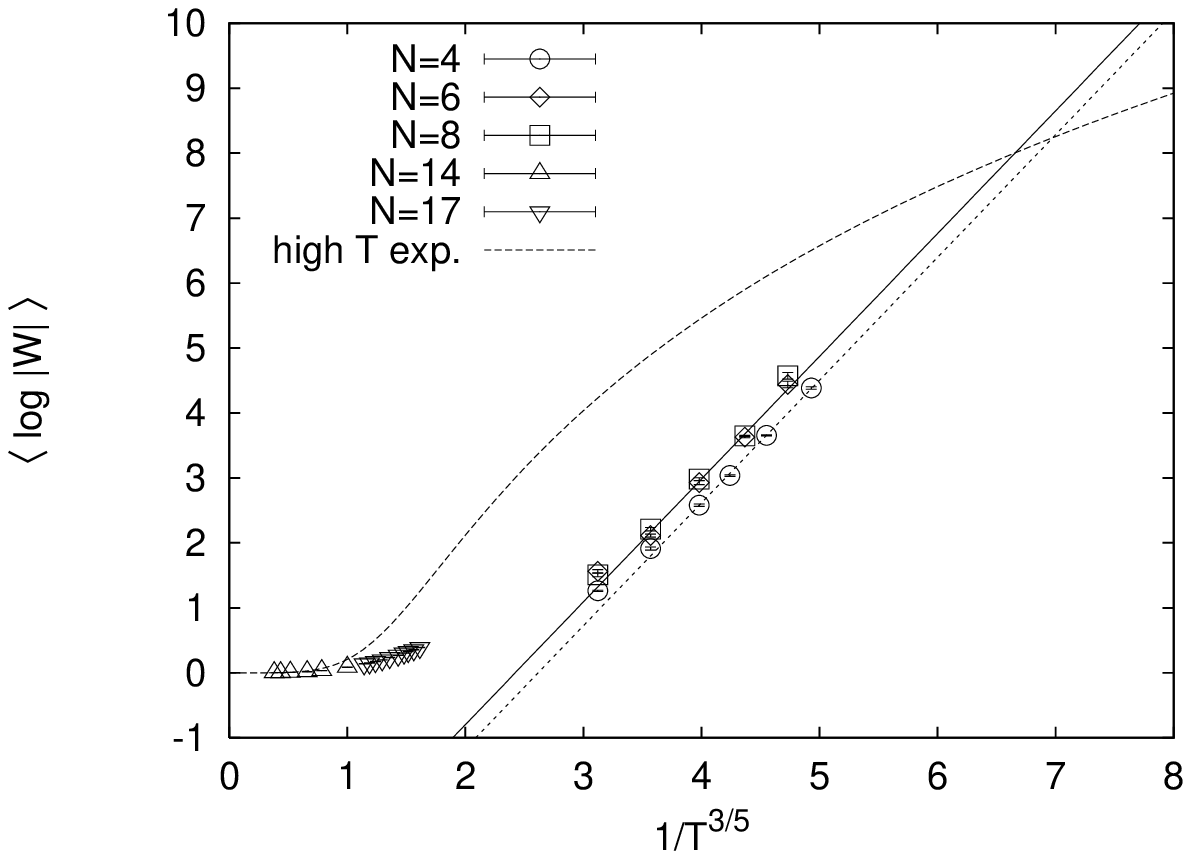}
%{bfss_continuum.eps}
\end{center}
\caption{
The plot of
%expectation value 
$\langle\log|W|\rangle$ for $\lambda=1$
%is plotted 
against $T^{-3/5}$. 
The cutoff $\Lambda$ is chosen as follows: 
$\Lambda=12$ for $N=4$; $\Lambda=0.6/T$ for $N=6,8$; 
$\Lambda=4$ for $N=14$; $\Lambda=6$ for $N=17$.
The dashed line represents the results
of the 
HTE
%high-temperature expansion
up to the next-leading order
%with extrapolations to $N=\infty$, 
for $N=14$, which are obtained by applying the method 
in Ref.\  \cite{HTE}.
The solid line and the dotted line represent fits
for $N=6$ and $N=4$ respectively,
to straight lines with the slope 1.89 
predicted from the gravity side at the leading order.
}
\label{fig:logW}
\end{figure}

As another prediction from the gauge-gravity duality, 
let us consider the Wilson loop, which winds around the temporal 
direction once, like the Polyakov line.
However, unlike the usual Polyakov line, 
we consider the one involving the adjoint scalar as
%it involves the adjoint scalar as
\beq
W \equiv \frac{1}{N} \,
\tr \,  {\cal P} \exp \left[ i \int_0^\beta
% dt \{ A(t) + i \,X_9(t) \} \right]
 dt \{ A(t) + i \, n_i \, X_i(t) \} \right]
%\sim  \exp \left( 
%\frac{ \beta R_{\rm Sch}}{2 \pi \alpha '}
%\right) 
\ ,
\label{Maldloop-def}
\eeq 
where $n_i$ is a unit vector in 9d, which can be chosen arbitrarily
due to the SO(9) invariance.
This object can be calculated on the gravity side by 
considering the minimal surface spanning the loop
in the dual geometry \cite{maldloop,PML}.
For the present model, the result is given by \cite{Hanada:2008gy}
\beq
\ln W = 
\frac{\beta R_{\rm Sch}}{2 \pi \alpha ' }
 = \kappa
 \left(\frac{T}{\lambda^{1/3}} \right)^{-3/5} \ ,
\label{Wilson-loop}
\eeq
where $R_{\rm Sch}$ is the Schwarzschild radius of the dual black hole
geometry and
\beq
\kappa = \frac{1}{2 \pi}\left\{ 
\frac{16 \sqrt{15} \pi^{7/2}}{7} 
\right\}^{2/5}  =  1.89\cdots \ .
\label{kappa-predicted}
\eeq
In figure \ref{fig:logW} 
we plot the log of the Wilson loop \cite{Hanada:2008gy} against
$T^{-3/5}$ anticipating (\ref{Wilson-loop}).
Indeed, at low temperature (to the right on the figure),
we find that the data points can be fitted nicely to a straight line
with a slope 1.89 in precise agreement with (\ref{kappa-predicted}).
The solid line corresponds to
%with the intercept.
$\langle \log |W| \rangle = 1.89 \, T^{-3/5} - 4.58 $,
where the existence of the constant term 
can be understood as $\alpha ' $ corrections.
This result demonstrates that one can extract the information of the dual 
geometry such as the Schwarzschild radius from the gauge invariant
observable (\ref{Maldloop-def}).
%, namely the Wilson loop.
It also confirms directly \cite{Hanada:2008gy}
the fuzz-ball picture \cite{Mathur:2008nj} of 
a black hole proposed to 
solve the information paradox.

%%%%% BFSS_continuum. %%%%%
%%%%%%%%%%%%%%%%%%

\begin{figure}[htb]
\begin{center}
\includegraphics[height=8cm,angle=270]{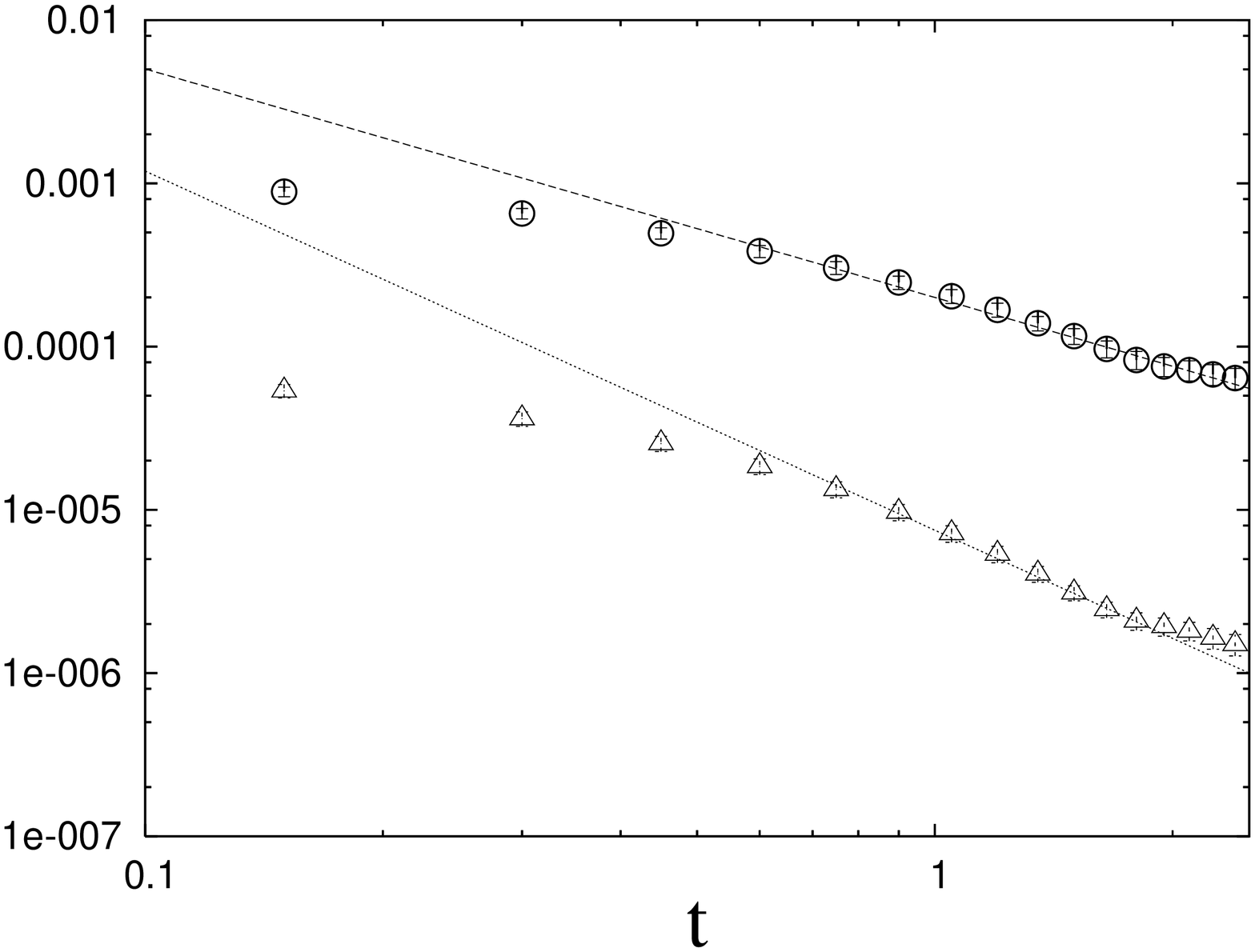}
%{bfss_continuum.eps}
\end{center}
\caption{
%The absolute value of
The two-point correlation function 
$\langle \mathcal{O}_{\ell}(t) \mathcal{O}_{\ell}(0) \rangle$
%of the operator
%(\protect\ref{defO})
is plotted for $\ell =4$ (circles)
%(inverted triangles) 
and $\ell =5$ 
%(diamonds) 
(triangles)
in the log-log scale.
Simulations were carried out at $N=3$, $\Lambda=12$, $T=0.2$.
The straight lines are fits to the predicted power-law behavior
(\protect\ref{SYpredict}).
}
\label{fig:Tpp}
\end{figure}

One can also predict various correlation functions from gravity.
This was done ten years ago by Sekino and Yoneya \cite{Sekino:1999av}
% obtained such a prediction
extending the Gubser-Klebanov-Polykov-Witten prescription
\cite{Gubser:1998bc}
%,Witten:1998qj}
to the present case.
For instance, let us consider an operator
\beq
\mathcal{O}_{\ell} = {\rm Str} (X_{i_1} X_{i_2} \cdots 
X_{i_\ell} ) \ ,
\label{defO}
\eeq
where the symbol ``S'' implies that all the indices are symmetrized.
The two-point correlation function of this operator
is predicted as
\beq
\langle \mathcal{O}_{\ell}(t) 
\mathcal{O}_{\ell}(0) \rangle
\sim \frac{1}{|t|^{p}} \ , 
\quad
p = \frac{4\ell - 9}{5} 
\label{SYpredict}
\eeq
at $\lambda ^{-1/3} \ll |t| \ll
\lambda ^{-1/3} N^{10/21}$.
In figure \ref{fig:Tpp} 
we plot the two-point correlation function
%for a different series of operators 
for $\ell = 4, 5$, which
agrees precisely with the predicted power law behavior.
See Ref.\ \cite{HNSY} for more details
% including
as well as results for other operators.

%%%%% BFSS_continuum. %%%%%
%%%%%%%%%%%%%%%%%%

\section{Extension to higher dimensions based on the large-$N$ reduction}
\label{sec:higherD}

%All the results presented in the previous section is nice, but
%the calculation 
In this section we discuss how one can extend the works
in the previous section to higher dimensions.
Respecting supersymmetry becomes more non-trivial 
in higher dimensions, but here again we stick to a non-lattice regularization.
For that purpose
we use the idea of the large-$N$ reduction, which we review briefly.
Let us consider U($N$) gauge theory on a $D$-dimensional torus,
and consider the Wilson loop defined by
\beq
W[C] = 
\left\langle \mathcal{P} {\rm exp} \left(i \int  A_\mu(x(\sigma)) 
\, \dot{x} ^\mu (\sigma) 
\, d\sigma
\right ) 
\right \rangle \ ,
\eeq
where the loop $C$ is specified by the embedding function
$C = \{ x^\mu (\sigma) \}$.
The corresponding 
large-$N$ reduced model can be obtained by simply reducing
the torus to a point.
This implies that we drop the $x$-dependence of the field $A_\mu (x)$
and obtain $A_\mu$.
The Wilson loop in the reduced model can be defined by
\beq
w[C] = 
\left\langle \mathcal{P} {\rm exp} \left(i \int  A_\mu
\, \dot{x} ^\mu (\sigma) 
% k ^\mu (\sigma) 
\, d\sigma
\right ) 
\right \rangle_{\rm red} \ .
\eeq
%where $k^\mu (\sigma)  = \dot{x}^\mu (\sigma)$.
Then the statement is that
\beq
\lim _{N\rightarrow \infty}w[C] 
= \lim_{N\rightarrow \infty}W[C] \ .
\eeq
The original idea was formulated on the lattice 
by Eguchi and Kawai \cite{Eguchi:1982nm}.
However, it was soon pointed out 
by Bhanot, Heller and Neuberger \cite{Bhanot:1982sh}
that there was
a problem due to the spontaneous breaking of the center symmetry,
which invalidates the proof of the statement.
%equivalence 
Several years ago Narayanan and Neuberger \cite{Narayanan:2003fc} proposed 
to avoid the spontaneous breaking of the center symmetry
by not reducing the torus
to a point completely, but keeping the volume finite in physical units.
More recently, 
Kovtun, \"Unsal and Yaffe pointed out \cite{Kovtun:2007py}
%it was pointed out \cite{Kovtun:2007py}
%Bringoltz and Sharpe pointed out 
that the original proposal with 
the one-site model actually works by adding an adjoint fermion 
if its mass is sufficiently small, which was supported by numerical
simulation \cite{Bringoltz:2009kb}.
See Bringoltz and Sharpe's contributions 
as well as Hietanen's one on this volume. 
(See also ref.\ \cite{Unsal:2008ch} for a proposal for
any non-abelian gauge theory.) 
%The scaling of the critical mass in the continuum
%limit is an important open question.

Here we use the idea of the large-$N$ reduction in order to study 
${\cal N}=4$ SYM on $R\times S^3$ as proposed by Ref.\ \cite{Ishii:2008ib}.
It actually differs from the original large-$N$ reduction
in that 
one deals with a curved space rather than a torus, which is a flat space.
The theory obtained after reducing the $S^3$ to a point is given
by the 1d SYM, which is nothing but the one discussed in the previous section,
plus some mass deformation, which 
preserves 16 supersymmetries of the undeformed
theory.
The additional terms are given by
\beq
\int dt \ 
\tr \left[ \,  \frac{1}{2}\mu^2 \sum _{i=1}^3 (X_i)^2 
+ \frac{1}{8}\mu^2 \sum_{a=4}^9 (X_a)^2 
+ i \mu \epsilon_{ijk} X_i X_j X_k 
+ \frac{3}{8} i \mu  \Psi \gamma_{123} \Psi \right] \ ,
\eeq
where $\mu$ is the deformation parameter, which is related to the
radius of the $S^3$ before reduction as 
\beq
R_{\rm S^3}=\frac{2}{\mu} \ .
\label{R-S3-def}
\eeq
This mass deformed theory possesses many classical vacua given by
$X_i = \mu L_i$,
where $L_i$ is an arbitrary (not necessarily irreducible) 
representation 
matrix of the SU(2) algebra
$[ L_i , L_j ] = i\,  \epsilon_{ijk}  \, L_k$.
These vacua preserve 16 supersymmetries, and they are all degenerate.

In order to retrieve the original 4d 
${\cal N}=4$ SYM, one has to pick up a particular vacuum
\beq
X_i = \mu \left(
\begin{array}{cccc}
L_i^{(n)} & ~ & ~ & ~ \\
~ & L_i^{(n+1)} & ~  & ~ \\
~ & ~ & \ddots & ~ \\
~ & ~ & ~ &  L_i^{(n+\nu-1)} 
\end{array}
\right)
\otimes {\bf 1}_k \ ,
\label{background}
\eeq
where $L_i^{(m)}$ represents the $m$-dimensional irreducible
representation of the SU(2) algebra.
Note, in particular, that there is an identity 
\beq
%\mbox{c.f.)~~~}
\sum_{i=1}^3 (L_i^{(m)})^2 = \frac{1}{4}
(m^2 -1)\,  {\bf 1}_m \ ,
\eeq
which implies that each of $L_i^{(m)}$ in
(\ref{background}) represents
a fuzzy sphere with the radius $\frac{\mu}{2} \sqrt{m^2-1}$.
%The configuration represents a sequence of fuzzy spheres with
%different radii.
In this construction one regards the $S^3$ as an $S^1$ fiber
on $S^2$, where the $S^2$ is represented by a fuzzy sphere, and
the $S^1$ fibration is represented by having many of them with
different radii.
A more detailed argument for the reduction is given 
in ref.\ \cite{Ishii:2008ib}.
The statement is that 
in the $k\rightarrow \infty $,
$n\rightarrow \infty $ and
$\nu \rightarrow \infty $ limits,
%% \footnote{Originally \cite{Ishii:2008ib},
%% it was stated that the $n \rightarrow \infty$ limit is also 
%% necessary, but some results in refs.\ \cite{Ishiki:2008te,Ishiki:2009sg}
%% suggested that one can also take $n=1$, for instance. 
%% This is an important issue from a practical point of view,
%% which deserves further investigations.
%% } 
one obtains  
the ${\cal N}=4$ U($\infty$) SYM on $R \times S^3$,
where the radius of $S^3$ is given by (\ref{R-S3-def})
and the 't Hooft coupling constant is given by
\beq
\lambda_{\rm SYM}
= 2 \pi^2 ( R_{\rm S^3})^3 \frac{g^2 k}{(2n+\nu-1)/2}  \ . 
\eeq
Note that one does not introduce the lattice structure 
anywhere in the formulation,
which is important for preserving supersymmetry.

\begin{figure}[htb]
\begin{center}
\includegraphics[height=6cm]{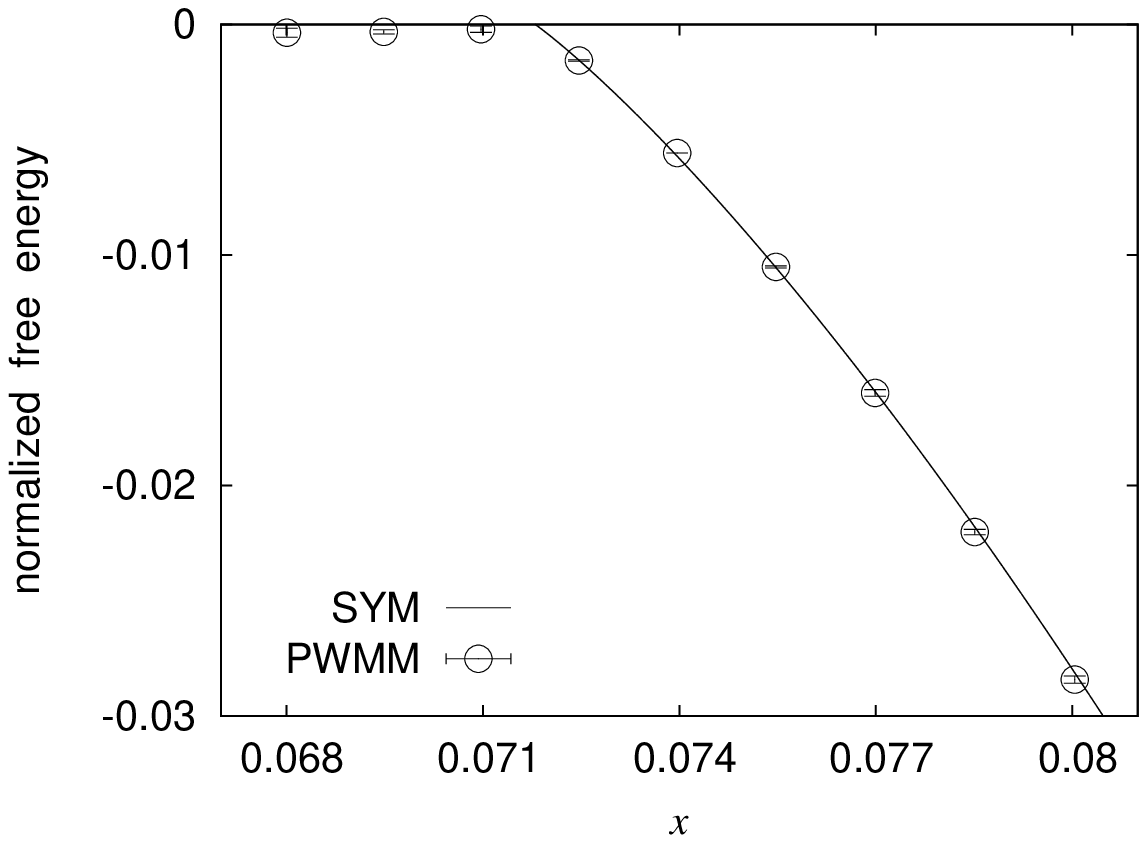}
%{bfss_continuum.eps}
\end{center}
\caption{
The normalized free energy
of the reduced model around the background 
(\protect\ref{background}) with $n=\frac{\nu+1}{2}$
in the $k\rightarrow \infty$ and $\nu\rightarrow \infty$ limits
is plotted against the dimensionless parameter 
$x=\exp (-\mu/2T)$ representing temperature 
near the critical point $x_c=0.072$.
The error bars represent the fitting error associated with
the extrapolation.
%The statistical errors
%are not shown because they
%are smaller than the symbol size.
The solid line represents the result \cite{Aharony:2003sx}
% (\ref{F_SYM})
for the ${\cal N}=4$ U($\infty$) SYM on $R \times S^3$.
}
\label{fig:free}
\end{figure}

A check of this novel large-$N$ reduction has been 
provided \cite{Ishiki:2008te}
%,Ishiki:2009sg}
in the weak coupling limit 
by studying
the deconfinement transition at finite temperature.
Figure \ref{fig:free} shows that the results obtained from the
reduced model with the background (\ref{background})
reproduce the known result \cite{Aharony:2003sx}
for ${\cal N}=4$ U($\infty$) SYM on $R\times S^3$
in the $k,n,\nu \rightarrow \infty $ limit.
%% $k\rightarrow \infty $ and
%% $\nu \rightarrow \infty $ limits.

%%%%% BFSS_continuum. %%%%%
%%%%%%%%%%%%%%%%%%

In order to test the approach at strong coupling, 
let us consider the circular Wilson loop in 
${\cal N}=4$ U($\infty$) SYM on $R^4$.
The expectation value of the Wilson loop is calculated
exactly for arbitrary coupling constant, and the result
is given as \cite{Erickson:2000af}
%is given as \cite{Erickson:2000af,Drukker:2000rr,Pestun:2007rz}
\beqa
\langle W_{\rm circular} \rangle
&=& \sqrt{\frac{2}{\lambda_{\rm SYM}}} \, 
I_1 (\sqrt{2\lambda_{\rm SYM}})
\label{WC-all-order}
\\
&\simeq &
\frac{e^{\sqrt{2\lambda_{\rm SYM}}}}
{(\frac{\pi}{2})^{1/2}(2\lambda_{\rm SYM})^{3/4}}
\quad \mbox{at $\lambda_{\rm SYM} \gg 1$}
\label{WC-at-strong}
\eeqa
in terms of the modified Bessel function of the first kind.
At strong coupling it agrees with the result obtained from the dual
geometry \cite{maldloop}.

Since the ${\cal N}=4$ SYM is conformally invariant,
the theory on $R^4$ is equivalent to
the theory on $R\times S^3$ through 
conformal mapping.
The circular Wilson loop on $R^4$
is mapped to a great circle on $S^3$ at a point
on $R$. 
(The size of the circular Wilson loop corresponds to the
position of the point on $R$, and the dilatation invariance on $R^4$
corresponds to the translational invariance on $R$.)
This Wilson loop can be represented in the large-$N$ reduced
model in a simple way as
\beq
W_{\rm circular} = \frac{1}{N} \, \tr \left[
\exp \left( 
i \frac{4\pi}{\mu} \{ X_3(t) + i \, X_4 (t)\}
\right) \right] \ ,
\eeq
where $t$ can be any value due to translational symmetry, and hence
one can take an average over it to increase statistics.

\begin{figure}[htb]
\begin{center}
\includegraphics[height=6cm]{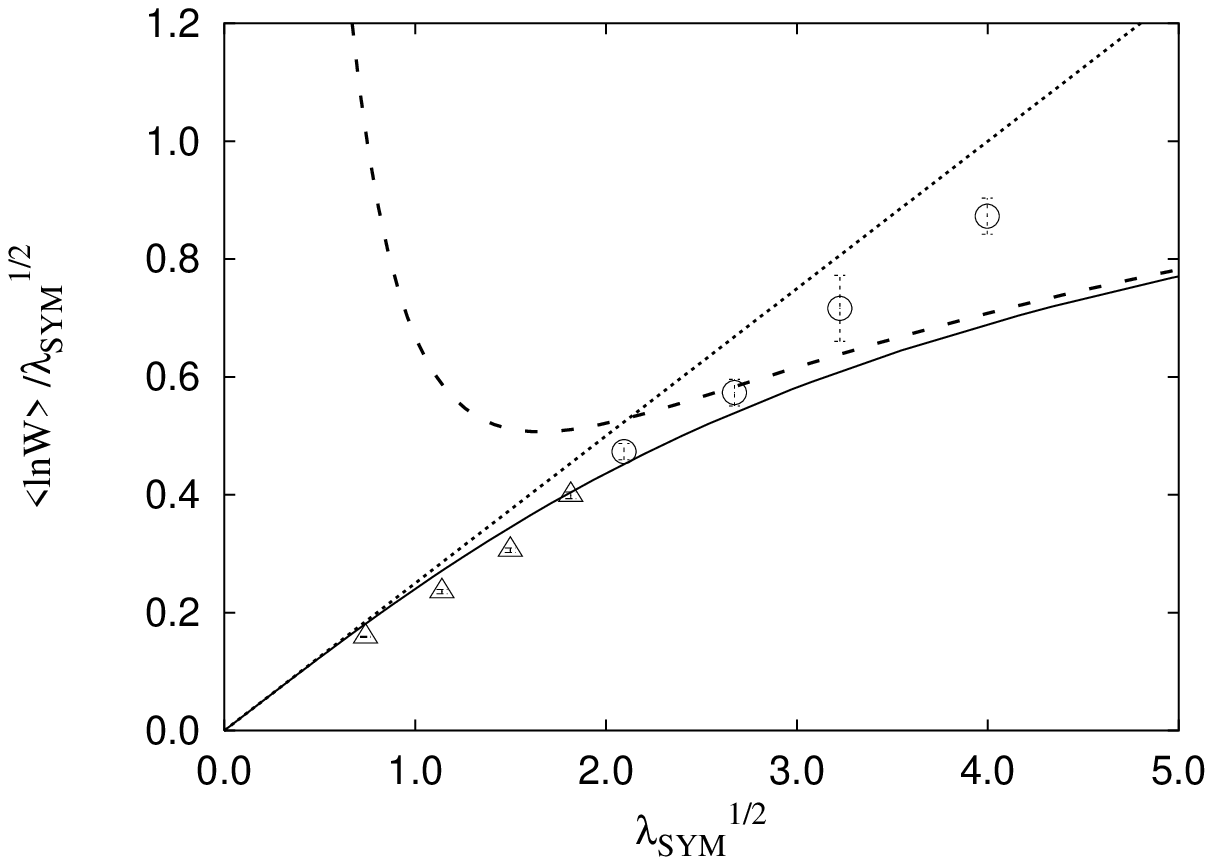}
%beijing3.ps}
%{bfss_continuum.eps}
\end{center}
\caption{
%The absolute value of
The log of the circular Wilson loop normalized by
$\sqrt{\lambda_{\rm SYM}}$ is plotted against 
$\sqrt{\lambda_{\rm SYM}}$.
We have performed the $\Lambda \rightarrow \infty$ extrapolation
(linear in $1/\Lambda$) using $\Lambda=6,8,10$.
The extent in the time direction is fixed to $\beta=5$.
The background is chosen to be $n=1$, $\nu=2$, $k=2$ for 
$\sqrt{\lambda_{\rm SYM}} < 2$, whereas
for $\sqrt{\lambda_{\rm SYM}} > 2$, we performed
an extrapolation to $k=\infty$ using the data for $k=2,3$
assuming that the finite-$k$ effects are O($1/k^2$).
The solid line represents the
all order result (\protect\ref{WC-all-order}).
The dashed line represents the behavior (\protect\ref{WC-at-strong})
at strong coupling, whereas
the dotted line represents the leading perturbative behavior
$\ln \langle W \rangle \simeq \frac{1}{4} \lambda_{\rm SYM}$.
% obtained from the localization technique.
}
\label{fig:wilsonSYM}
\end{figure}

In figure \ref{fig:wilsonSYM} we present
our preliminary results for the circular Wilson loop \cite{IHNT}.
(Here after, we present results obtained by imposing 
periodic boundary conditions on fermions,
since we are interested in zero temperature.)
We also plot the all order result (\ref{WC-all-order}).
%% In order to see which regime is weak coupling and which regime 
%% is strong coupling, we expand the all order result
%% around $\lambda_{\rm SYM}=0$ and $\lambda_{\rm SYM}=\infty$,
%% and plot the results up to the subleading terms.
Although the matrix size
%chosen background 
is obviously too small,
our Monte Carlo results look promising.
Note, in particular, that we already start to observe
a bent from the weak coupling behavior towards
the strong coupling behavior.

%%%%% BFSS_continuum. %%%%%
%%%%%%%%%%%%%%%%%%

Next let us consider chiral primary operators
such as $\tr Z^J$, where
$Z=\frac{1}{\sqrt{2}}(X_4 + i \, X_5)$.
%\label{chiral-pr-op}
The two-point function can be calculated
in the weak coupling limit of $\mathcal{N}=4$ U($\infty$) SYM on $R^4$ as
%ia free theory as
\beq
\langle \tr Z^J (z_1) \, 
\tr Z^{\dag J} (z_2) \rangle_{R^4}
= \frac{c_J }{|z_1 -z_2|^{2J}}  \ , \quad \quad
c_J= J \left(\frac{\lambda_{\rm SYM} }
{4 \pi^2} \right)^J \ .
\label{free-r4}
\eeq
It is known in $\mathcal{N}=4$ SYM that 
the supersymmetry 
non-renormalization theorem holds for the two-point 
functions. Hence, the result (\ref{free-r4})
actually holds for arbitrary coupling constant.
As in the case of the circular Wilson loop,
one can make a conformal mapping to $R \times S^3$,
and obtain for $J=2$, for instance,
\beq
\int d \Omega_3 \int d \Omega_3 '
\ \langle \tr Z^2 (t , \Omega_3) \, 
\tr Z^{\dag 2} (0 , \Omega_3 ' ) \rangle_{R\times S^3}
= \frac{c_2 e^{-\mu t}}{1-e^{-\mu t}}   \ ,
\label{2pt-N4}
\eeq
where we have integrated over the $S^3$
since the operator 
$\tr Z^{J}(t)$ in the reduced model corresponds to
$\int d \Omega_3 \, \tr  Z^{J}(t,\Omega_3)$
in the SYM on $R \times S^3$.

\begin{figure}[htb]
\begin{center}
\includegraphics[height=6cm]{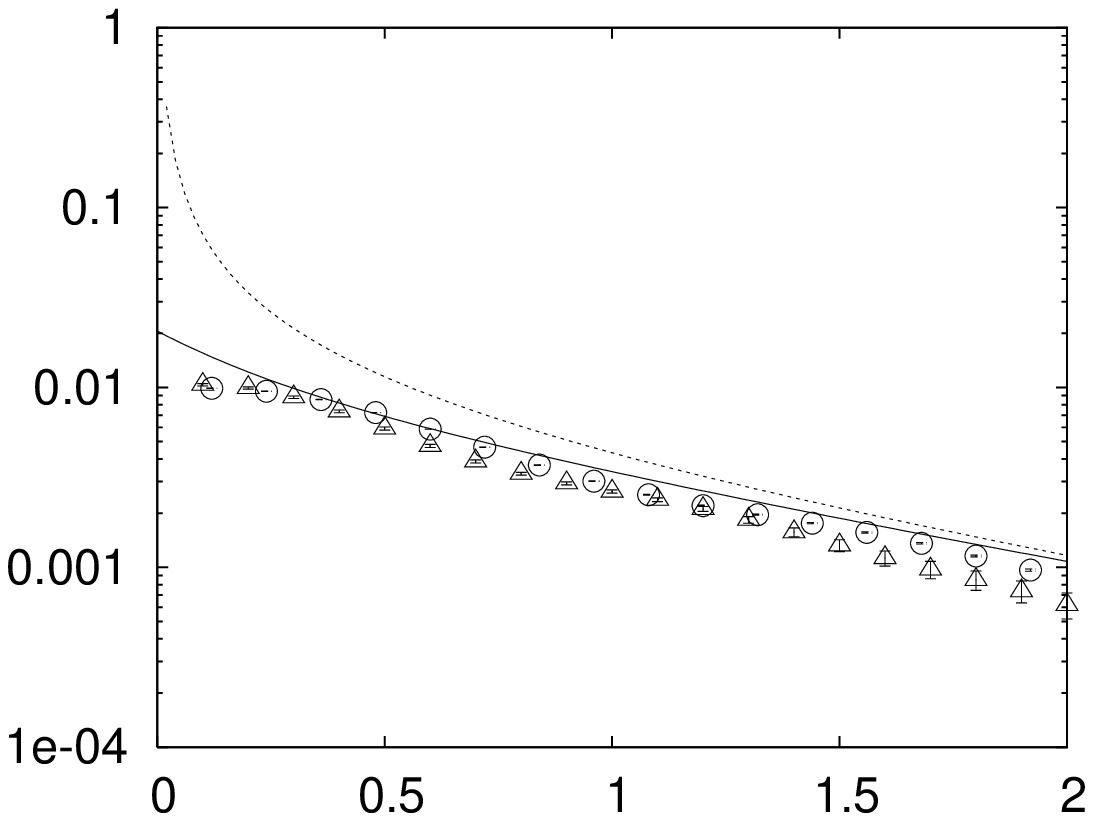}
%{bfss_continuum.eps}
\end{center}
\caption{
%The absolute value of
The two-point function of the chiral primary operator
%(\protect\ref{chiral-pr-op})
is calculated from the reduced model
for a fixed background (\protect\ref{background})
with $n=3$, $\nu=2$ and $k=2$.
%with only $L_i^{(3)}$ and $L_i^{(4)}$ as the diagonal blocks
%and with $k=2$ as a preliminary study. 
The UV cutoff is chosen to be $\Lambda=10$.
The circles represent results for $\mu=3$, $\beta=4$,
which corresponds to $\lambda_{\rm SYM}=0.24$.
The triangles represent results for $\mu=1$, $\beta=10$,
which corresponds to $\lambda_{\rm SYM}=6.4$.
Surprising, the results lie more or less on top of each other.
%We also plot the result for the ${\cal N}=4$ SYM.
The solid line represents the analytic 
result in the weak coupling limit of the reduced model
for the same background in the $\Lambda \rightarrow \infty$
and $\beta \rightarrow \infty$ limits, which shows reasonable
agreement with the Monte Carlo data. 
The dotted line represents the analytic result
in the weak coupling limit of 
the $\mathcal{N}=4$ U($\infty$) SYM,
which is expected to be reproduced from the reduced model
in the $k,n,\nu \rightarrow \infty$ limit.
%% in the $\nu \rightarrow \infty$ and
%% $k \rightarrow \infty$ limits.
}
\label{fig:2pt}
\end{figure}

In figure \ref{fig:2pt} we plot our preliminary results
for $\langle \tr Z^2 (t) \, 
\tr Z^{\dag 2}(0) \rangle_{\rm red}$
against the dimensionless time $\mu t$ \cite{IHKNT}.
Surprisingly our results for two different values of 
$\lambda_{\rm SYM}$ turn out to be very close to each
other.
% if we rescale the time coordinate appropriately.
This suggests that the non-renormalization theorem
actually holds
for each background.
In fact we can obtain results 
in the weak coupling limit of the reduced model
%in free theory
for the background (\ref{background}),
and show that the correlation function approaches the one
(\ref{2pt-N4}) for the ${\cal N}=4$ U($\infty$) SYM 
in the $k , n , \nu \rightarrow \infty$ 
limit \cite{IHKNT}.
%% in the $k\rightarrow \infty$ 
%% and $\nu \rightarrow \infty$ limits \cite{IHKNT}.
Therefore, if we are able to confirm the 
non-renormalization theorem for each background,
it immediately implies that we can reproduce
the results for the ${\cal N}=4$ U($\infty$) SYM 
from the reduced model
at arbitrary 't Hooft coupling constant.
%also at strong coupling.

%%%%% BFSS_continuum. %%%%%
%%%%%%%%%%%%%%%%%%

What we have presented so far should be considered as 
a check of our method.
More interesting quantities are those which are \emph{not}
obtained in the strongly coupled gauge theory,
and yet there exist interesting predictions from gravity.
Calculating such quantities by our method will clearly
provide a new test of the AdS/CFT correspondence.

For instance, we can consider the rectangular Wilson loop
in $R^4$, which behaves as
\beq
\langle W(T\times R) \rangle
= \exp \left(  \frac{ \gamma \, T}{R}
\right)
\eeq
at $T \gg R$ due to conformal symmetry.
This is in striking contrast to the area law in 
pure Yang-Mills theory.
%confining theories.
%% This implies that the static potential between
%% a quark and an anti-quark is given by
%% \beq
%% V(R) = - \frac{c}{R} \ .
%% \eeq
In particular, the AdS/CFT correspondence predicts
\beq
\gamma = \frac{4 \pi^2 \sqrt{2\lambda_{\rm SYM}}}
{\Gamma^4(1/4)}
\eeq
at strong coupling \cite{maldloop}.

It would be also interesting to study higher point functions of
the chiral primary operators.
(See ref.\ \cite{Berenstein:2008jn}
for calculation of extremal 3-point functions
by simulating a ``truncated theory'' composed 
of six commuting bosonic matrices.)
In particular, AdS/CFT predicts that 
%The AdS/CFT predicts that 
non-extremal 4-point functions 
violate the non-renormalization theorem.
It is interesting to check whether this is indeed the case,
and to obtain explicit results, which can be compared with the
prediction from gravity.

\section{Summary} 
\label{sec:summary}

I hope I have convinced the readers that 
non-lattice simulations are indeed useful for
studying supersymmetric large-$N$ gauge theories
in the strongly coupled regime.
%I have discussed non-lattice simulations of
%supersymmetric large-$N$ gauge theories.
%In particular, 
%
%% I presented numerical results for 
%% models with maximal supersymmetries,
%% which revealed direct connections to gravity as predicted 
%% by the gauge-gravity duality.
In the first part, I discussed
% the results for
the 1d SYM with 16 supercharges, which reproduced
% was studied, and the
black hole thermodynamics and the Schwarzschild radius 
of the dual geometry.
% were reproduced.
These results revealed direct connections to gravity as predicted 
by the gauge-gravity duality.
In particular, the gauge theory results 
provided a clear understanding of the
microscopic origin of the black hole thermodynamics.
In the second part, I discussed how one can extend
these works to higher dimensions
%can be extended to higher dimensions
%We can also study higher dimensional gauge theories 
by using the novel large-$N$ reduction.
%, which preserves 16 supersymmetries.
I have presented some preliminary results 
for ${\cal N}=4$ U($\infty$) SYM on $R\times S^3$.
This theory is superconformal, and it actually has 32 supersymmetries.
Our formulation preserves 16 supersymmetries 
in the $\Lambda \rightarrow \infty$ limit, 
and the remaining half of 
the supersymmetries are expected to be 
restored without fine-tuning by increasing the matrix size.
%in the continuum limit 
%
%% This seems to be the case at weak coupling as we have seen 
%% by explicit calculations,
%% but whether it holds also at strong coupling has to be checked.
%
%% Finally, let us comment that one can also study ${\cal N}=8$ SYM
%% on $R \times S^2$, and also theories with less supersymmetries, which
%% are technically less demanding.
%% However, the large-$N$ reduction does not work in the bosonic case,
%% due to the instability of the background (\ref{background}).

In a way, what we have seen is 
the beginning of a whole new field of research
%as the lattice gauge theory was in the early 80s.
analogous to the situation of the lattice gauge theory in early 80s.
%just the tip of an iceberg.
Now with the aid of supersymmetry and large $N$, we have just started to
explore superstring theory and quantum space-time from first principles. 
%I believe that Monte Carlo simulation will play a major role
%in exploring supersymmetry, superstrings and quantum space-time
%in the next decade.

%\acknowledgments

%% I would like to thank M.\ Hanada, M.\ Honda, G.\ Ishiki
%% and A.\ Tsuchiya for reading the manuscript carefully.
%% Most of the calculations were carried out on PC clusters at KEK.
%% This work is supported 
%% %in part 
%% by Grant-in-Aid for Scientific
%% Research (Nos.\ 19340066 and 20540286)
%% from Japan Society for the Promotion of Science.
%% %the Ministry of Education, Culture, Sports, Science and Technology.

\end{document}